\documentclass[aps,twocolumn]{revtex4}
\usepackage{graphicx}
\usepackage{booktabs}
\usepackage{xcolor}
\usepackage{hyperref}
\usepackage{multirow}
\usepackage{setspace}

\newcommand{\comment}[1]{}

\begin{document}

\title{From Big Data To Important Information} 
\date{April 1, 2016}    
\author{Yaneer Bar-Yam}
\affiliation{New England Complex Systems Institute \\ 
210 Broadway, Suite 101, Cambridge, MA 02139, USA}

\begin{abstract}
Advances in science are being sought in newly available opportunities to collect massive quantities of data about complex systems. While key advances are being made in detailed mapping of systems, how to relate this data to solving many of the challenges facing humanity is unclear. The questions we often wish to address require identifying the impact of interventions on the system and that impact is not apparent in the detailed data that is available.  Here we review key concepts and motivate a general framework for building larger scale views of complex systems and for characterizing the importance of information in physical, biological and social systems. We provide examples of its application to evolutionary biology with relevance to ecology, biodiversity, pandemics, and human lifespan, and in the context of social systems with relevance to ethnic violence, global food prices, and stock market panic. Framing scientific inquiry as an effort to determine what is important and unimportant is a means for advancing our understanding and addressing many practical concerns, such as economic development or treating disease. 
\end{abstract}

\maketitle

\section{Overview}

Changing disease to health and economic instability to growth are among the complex challenges we face today. How can we turn the massive quantities of data that are increasingly available towards addressing these pressing problems? The data provide abundant detail, but generally carry no labels for guidance about which pieces of information are important for determining successful interventions. The questions we need to address are about properties of complex systems---human physiology, the global economy. Addressing questions about such systems requires disentangling the intricate dependencies and multiple causes and effects of behaviors, and recognizing that behaviors range across scales from microscopic to macroscopic. 

Here we argue that the key to addressing these questions is to focus on the way behavior at different scales are related, and how dependencies within a system lead to the large scale patterns of behavior that can be characterized directly without mapping all of the intricate details. The approach builds upon an understanding of how to aggregate component behaviors to identify larger scale behaviors, an approach developed in the ``renormalization group'' study of phase transitions in physics, and generalized here to multiscale information theory. In this framework, information itself has scale and larger scale information is the most important information to know, with progressively finer scale information only of importance to provide detail when necessary. This analysis focuses attention on information characterizing how to affect the largest scale behaviors of the system. The method is a shortcut for the infeasible effort of mapping all of the causes and effects that extend from molecular to global scales of biological and social systems. Specific causes and effects that need to be studied are only a few compared to the many that underly the system behavior at all scales. It is thus a tremendous simplification compared to extending traditional approaches to the desired outcome. On the other hand, the resulting formalism is challenging to execute in particular circumstances. When properly applied, the result is clear guidance about how to intervene and solve major problems, a result justifying the high level of directed effort involved. Successful examples demonstrate it is possible to apply it to a wide variety of scientific questions and real world problems, though many aspects of how to proceed more generally remain to be developed. Where the effects of interest range across scales, other methods must be applied. Both opportunities and limitations of the method can be understood within the general formalism that describes the approach. 

The approach is complementary to many other strategies that are usefully applied to complex systems, including not just big data but also network models, agent based models, game theory, system dynamics, machine learning, stochastic modeling, coupled differential equations, and other frameworks whose starting point is a particular representational framework. It is closer in spirit to fractals and chaos in their attention to the role of scales, but as with the other approaches, it does not adopt their specific representational strategies. In the approach described here the strategy is to describe the largest scale behaviors with a minimal but faithful representation, and each of the different representational strategies, or combinations of them, may be used as appropriate. 

This paper reviews and extends basic approaches to the quantitative analysis of systems. While many of the concepts have been developed in physics, the generalization provided here can be better applied across complex biological and social systems. The paper is written to be accessible to a wide audience and yet to provide essential insights that are important for physicists and mathematicians that are interested in expanding the quantitative understanding of complex systems. Despite the effort to make it accessible, it is not a general tutorial in complex systems science concepts and methods, for which other resources are available \cite{dcs,sayamatext}. Reviewing certain basic concepts is necessary here as they are not typically laid out in a way that is conducive to generalization. Examples are also included of applications to biological and social systems. While they do not manifest the full power of the fundamental approach, each illustrates aspects of its application to a highly complex system with real world importance, and without reference to the fundamental, general approach, it would be unclear how they are being studied. This paper is not a ``how to'' but rather a conceptual framing and motivation, so that there is much room left for other papers to provide practical guides for future work.

In Section II, we review the essential nature of traditional scientific approaches based upon calculus and statistics that rely upon separation of scales of behavior. We point out why they are not effective for dependencies of complex systems that generate multiscale patterns of behavior. In Section III, we describe how an essential insight about the breakdown of calculus and statistics arose from the study of phase transitions in materials leading to the development of concepts of renormalization group that uses a multiscale approach. In Section IV, we frame a generalization of this approach to complex systems by introducing the complexity profile, a multiscale characterization of complex systems, and the concept of descriptions that are reliable at a particular scale of observation. Section V discusses the concept of universality, formalized in renormalization group, which serves in the more general complex systems context as a way to recognize the importance of characterizing the behavior of systems at large scale. Section VI briefly addresses complications that arise due to dynamics of amplification, dissipation and chaos which do not preserve scale over time. Section VII provides a brief discussion of several examples, population biology, ethnic violence, the dynamics of food prices, market crashes, organizational structures and references to others. Finally in Section VIII, we point out that without a focus on universality, there can be no effective understanding of systems as each individual observation is of a different microscopic state. Thus, anchoring scientific inquiry in a scale sensitive approach focused on large scale behaviors is essential. Without attention to scale, any approach is destined to miss essential large scale properties and devote most of its attention to irrelevant fine scale details. 

\section{separation of scales}

One of the central insights about complex systems is that the effect of dependencies among components cannot be fully represented by traditional mathematical and conceptual approaches. A key to their limitation is that they are applicable only to systems in which there is a separation of behavior between the micro and macro scales. Interactions among the parts that cause behaviors across scales violate this separation. 

\begin{figure}[tb!]
\centering \includegraphics[width=3in]{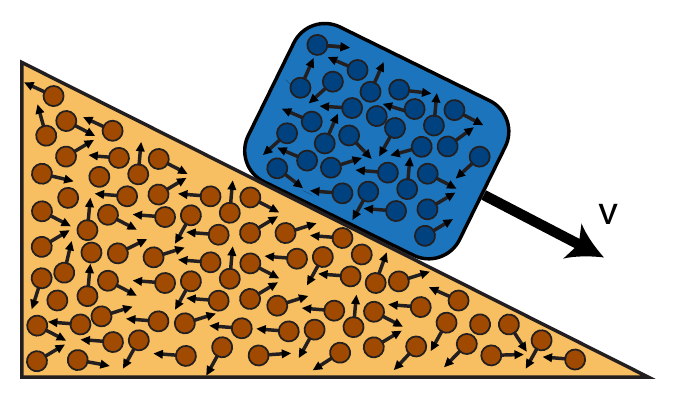}
\caption{\label{fig2} Schematic diagram of a block (with a velocity at a particular moment, $v$) sliding down an inclined plane. The macroscopic motion subject to gravity and friction may be treated using Newton's laws of motion, while the microscopic behavior of the atoms may be treated using thermodynamics by considering the local oscillations of groups of atoms as random and independent (the probability that one group is in a particular state is independent of the state of another group); the statistical treatment of that movement leads to the determination of pressure and temperature of the block and the inclined plane.}
\end{figure}

Consider a block sliding down an inclined plane. In a traditional approach, micro and macro scales are treated separately. To address dynamics at the micro scale---the molecules---we average over them and, using thermodynamics, describe their temperature and pressure. To address dynamics at the macro scale---the motion of the block on the inclined plane---we use Newtonian physics to talk about their large scale motion (see Fig. \ref{fig2}). In this case, the pieces can be considered to be acting either independently, like the random relative motion on the micro scale, or coherently, like the average motion on the macro scale. Since the scales are sufficiently distinct, separated by orders of magnitude, we do not encounter a problem in describing them separately. Finally, often unstated, the structures of the block and the plane are considered fixed. 

Thus, traditionally, there were three aspects of a system: fine scale, dynamic, and fixed. A glass of water on a table with an ice cube in it might be treated by considering the movement and melting of the ice cube, the average over molecular vibrations, and the fixed structure of the glass. At longer time scales, the water will evaporate, the glass will flow, the table may rot, but this is not important at a particular scale (or a range of scales) of observation. 

Consider the earth viewed from space. The earth is highly complex. Still, we can describe it as a planet orbiting the sun in a predictable fashion. Most of the details of what happens on Earth play no role at the scale of its orbit. For the earth, at the orbital scale, all the internal structure can be averaged to a point. The bodies of the solar system are assumed unchanging and the material of each of them is separated from other solar objects. The dynamic behavior can then be modeled and predicted. 

When separation of scales works, we can describe not only the system as it exists in isolation, but also how it responds to external forces. Forces that act on the earth at the scale of orbital motion couple to the dynamic behavior that occurs at that scale. Thus if we were to consider a new celestial body entering the solar system, unless it disrupted the structure of the system (i.e. by shattering a planet) and as long as we continue to be interested in the scale of orbital motion, we can describe the behavior of the system using these same degrees of freedom. 

For complex systems, it is still true that the questions we most want to answer have to do with the larger scale information. Significantly, the scale of description and scale of interactions are similar. When we have a description of the larger scale behavior we are also considering the larger scale impacts of the environment on the system and reciprocally. 

But many systems, especially those we are interested in understanding and influencing, are not well described by separate micro and macro scales. Consider a flock of birds. If all of the birds flew independently in different directions, we would need to describe each one separately. If they instead all went in the same direction, we could simply describe their average motion. However, if we are interested in their movement as a flock, describing each bird's motion would be too much information and describing the average would be too little information. Understanding complex behavior that is neither independent nor coherent is best described across scales. This requires knowing which information can be observed at a scale of interest. A generalized framing of this example can be applied to a wide range of complex systems.

\section{Theory-experiment contradiction and multiscale insight}

The key ideas can be understood from the difference between traditional physics and an approach developed in statistical physics beginning in the 1970s anchored in the method of renormalization group. Modeling in this framework allows distinguishing what can be observed at the largest scale. To explain the concepts of this formalism, we describe its development in the study of materials \cite{domb,a,rg,KWSciAm,kadanoff,nelson,kardarbook}. We then describe a generalization to complex biological and social systems.

Central to the study of matter is that movements of individual atoms are not visible to us. Instead, we use pressure, temperature and volume to describe both what we see and how we can manipulate matter using forces. For example, a piston compressing a gas reduces the volume and increases the pressure, and heat transfer to a material causes its temperature to rise. The key concept underlying our ability to make such descriptions is scale: The fine scale (microscopic) behaviors of atoms are not important to an observer or to their manipulation of a system; and the large scale (macroscopic) properties we observe and manipulate reflect average or aggregate properties of atomic motion. 

This approach was formalized in the 1800s through statistical physics. It appeared to solve the problem of determining properties of a material in equilibrium by minimizing the free energy relative to the macroscopic variables. This almost always works. However, in the study of phase transitions, e.g. between water and steam or between ferromagnet and paramagnet, properties were found not to be correctly given by this method for special conditions called second order phase transition points. This phenomenon proves to be a relatively simple illustration of a complex system, where the elements act neither fully independently nor fully coherently, and the separation of scales breaks down.

\begin{figure}[tb!]
\centering \includegraphics[width=3.0 in]{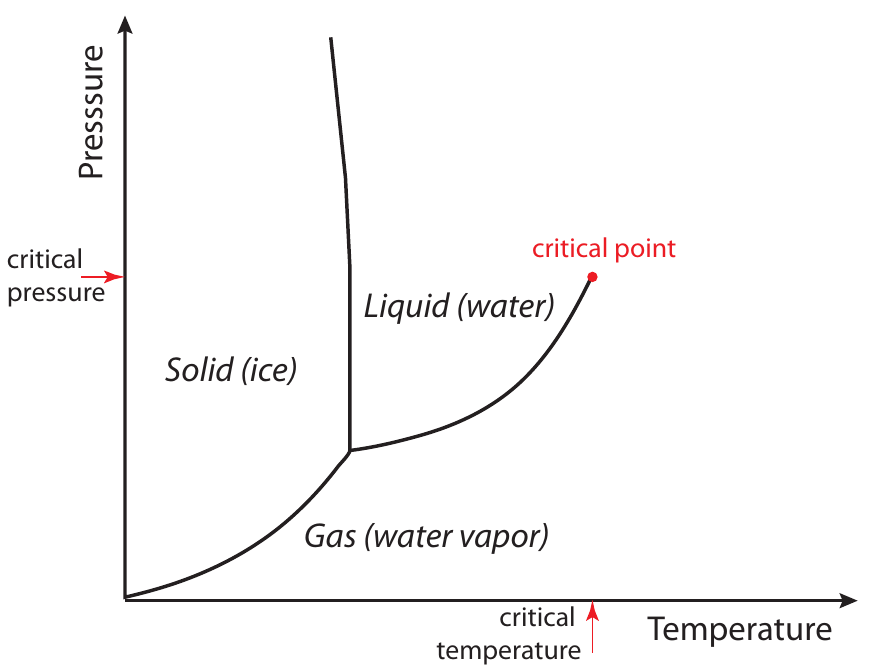}
\caption{\label{fig3} The phase diagram of water. The line of transitions between liquid water and water vapor stops at the critical point (red dot). At that point the fluctuations between liquid-like and vapor-like densities extend across the system so that the system is not smooth (violating the assumptions of calculus) and averages are not well behaved (violating the assumptions of statistics). A new method that considers behaviors across scales, renormalization group, was developed to address this and similar questions. }
\end{figure}

Consider the transition between water and steam. At a particular pressure we can cause a transition between water and steam by raising the temperature. At the transition temperature the density changes abruptly---discontinuously. As we raise the pressure, we compress the steam and the change in density at the transition temperature decreases (see Fig \ref{fig3}). There is a point where the transition stops, and there is no longer a distinction between water and vapor. This end point is called a second order transition point, at the end of the first order transition line. Near this point, the discontinuity of the density between liquid and gas phases becomes zero (hence the term second order transition). The way it does so has the form of a power law $\rho \propto x ^ \beta$, where $x$ is the distance along the transition line from the second order transition point. There are many other materials that have phase transitions lines that end at points, called second order phase transitions, or critical points. Power laws are ubiquitous near critical points. The exponent that was found empirically was $\beta \approx 0.33$. The same value of the exponent is found in many cases, including at critical points in both magnets and liquids \cite{exponents,exponents2,exponents3}. 
However, the theoretical prediction based upon free energy minimization is found to be $0.5$ \cite{landau}. The derivation starts from an analytic expansion of the free energy in the density around the critical point, then setting its derivative to zero to obtain the minimum (Landau theory). 

This surprising discrepancy between observations and theory compelled a dramatic change in our understanding. Our usual methods of calculus and statistics fail at this point because their assumptions no longer hold true. Calculus assumes that matter is smooth and statistics assumes that averages over large numbers of objects are well defined. Away from the critical point these assumptions are justified, since the microscopic behavior of atoms is well separated from the macroscopic behavior of the material as a whole. Different parts of the material appear essentially the same, making it smooth, and any (local) average over atomic properties has a single well defined number. However, at the critical point, the density fluctuates---between water-like and vapor-like conditions---so that the material is not smooth and the average 
taken of the material as a whole is not representative of the density at any particular location or time.
Near the critical point, the matter is composed of patches of lower and higher density, and 
this patchiness occurs on all scales, even at the macroscopic scale. 

In order to mathematically solve this problem, the renormalization group was developed \cite{rg}. In the renormalization group method, we consider the system at multiple scales (levels of resolution). The spatially varying macroscopic density at one level of resolution is related to that at a larger scale by performing local averages rather than a global average. This averaging relates the free energy at one scale of observation to the free energy at a larger scale. The properties of the system can be found from how the behavior varies with scale, taking the limit of arbitrarily large (infinite) scales. The mathematics is not easy, but it yields exponents that agree with the phenomenology \cite{rg,a,exponents}.  Since its development, renormalization methods have enabled many advances in addressing questions about the structure and dynamics of materials \cite{degennes,kardar}.  

The reason that different results were obtained is that the free energy in this case is not just a function of the average density. Still, it is not necessary to consider interactions among individual atoms. For a liquid undergoing transition to a vapor, the free energy depends on the spatial variation of the density, i.e. how the local densities at different locations interact with each other. There are many possible interactions between local densities that could contribute to the free energy. However, only some of them are important. The renormalization group is a method for determining which parameters describing the interaction are important and which are not. ``Relevant'' parameters are those parameters of the free energy that increase with scale; ``irrelevant'' parameters are those that decrease with scale. Because there are so many atoms in matter, the irrelevant parameters cannot affect our observation. We can consider only the relevant parameters. We might measure irrelevant parameters microscopically, but they won't affect macroscopic changes in the material or our interactions with it near the critical point.

\section{Representations and Information as a function of scale}

Why does Landau theory, based upon calculus and statistics, fail? The problem is not in the free energy minimization, the problem is in the representation of the material. Assuming that the free energy can be written as a function of the average density does not work. A single variable does not capture what is going on in the material. On the other hand, the positions of all of the atoms are not necessary. The spatial pattern of the densities is what we need. Moreover, the closer we look, the more values of the density we need. {\it The way we represent the system is the crux of the matter}---the sufficiency of the representation at the scale of observation we choose.

A representation is a map of the system onto mathematical variables. More correctly, a representation should be understood as a map of the set of possible states of the system onto the possible states of the mathematical variables. A faithful representation must have the same number of states as the system it is representing. This enables the states of the representation to be mapped one to one to the states of the system. If a model has fewer states than the system, then it can't represent everything that is happening in the system. If a model has more states, then it is representing things that can't happen in the system. Conventional models often do not take this into account and this results in a mismatch of the system and the model; they are unfaithful representations and do not properly identify the behavior of the system, and thus ultimately its response to environmental forces or interventions we might consider. Because we are interested in influencing the system, we only want to know the distinctions that matter. We have to focus attention on those states that are distinguishable at a particular scale of observation. 

\begin{figure}[tb!]
\centering \includegraphics[width=2.8 in]{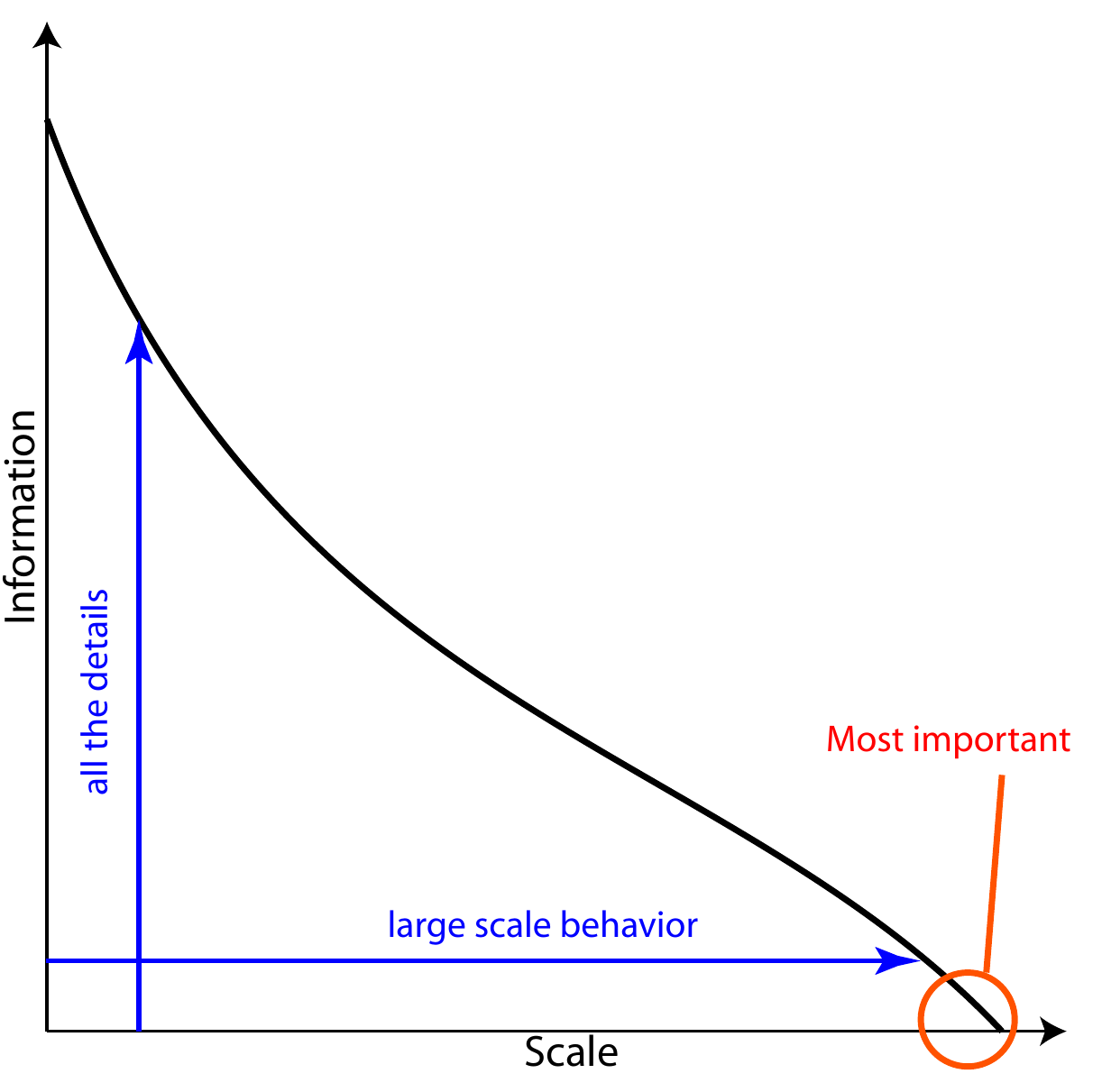}
\caption{\label{fig4}  The complexity profile is the amount of information that is required to describe a system as a function of the scale of description. Typically, larger scales require fewer details and therefore smaller amounts of information. The most important information about a system for informing action on that system is the behavior at the largest scale.}
\end{figure}

To formalize these ideas for complex systems, it is useful to understand information as related to scale. We define the complexity profile \cite{dcs,multiscalevariety} as the amount of information necessary to represent a system as a function of scale. Information theory defines the amount of information in a message as the logarithm (base 2) of the number of possibilities of the message---the number of bits needed to represent the set of possible messages. Thus, the complexity profile is given by the number of possible states of the system at a particular scale. Typically, the finer the scale of inquiry about a system, the more information is needed to describe it (Fig. \ref{fig4}). The complexity at the finest scales is finite because of quantum uncertainty and is equal to a universal constant, $1/k_B \ln(2)$, times the entropy for a system in equilibrium, where $k_B$ is Boltzmann's constant \cite{multiscaleACS}.

\begin{figure}[tb!]
\centering \includegraphics[width=1.5in]{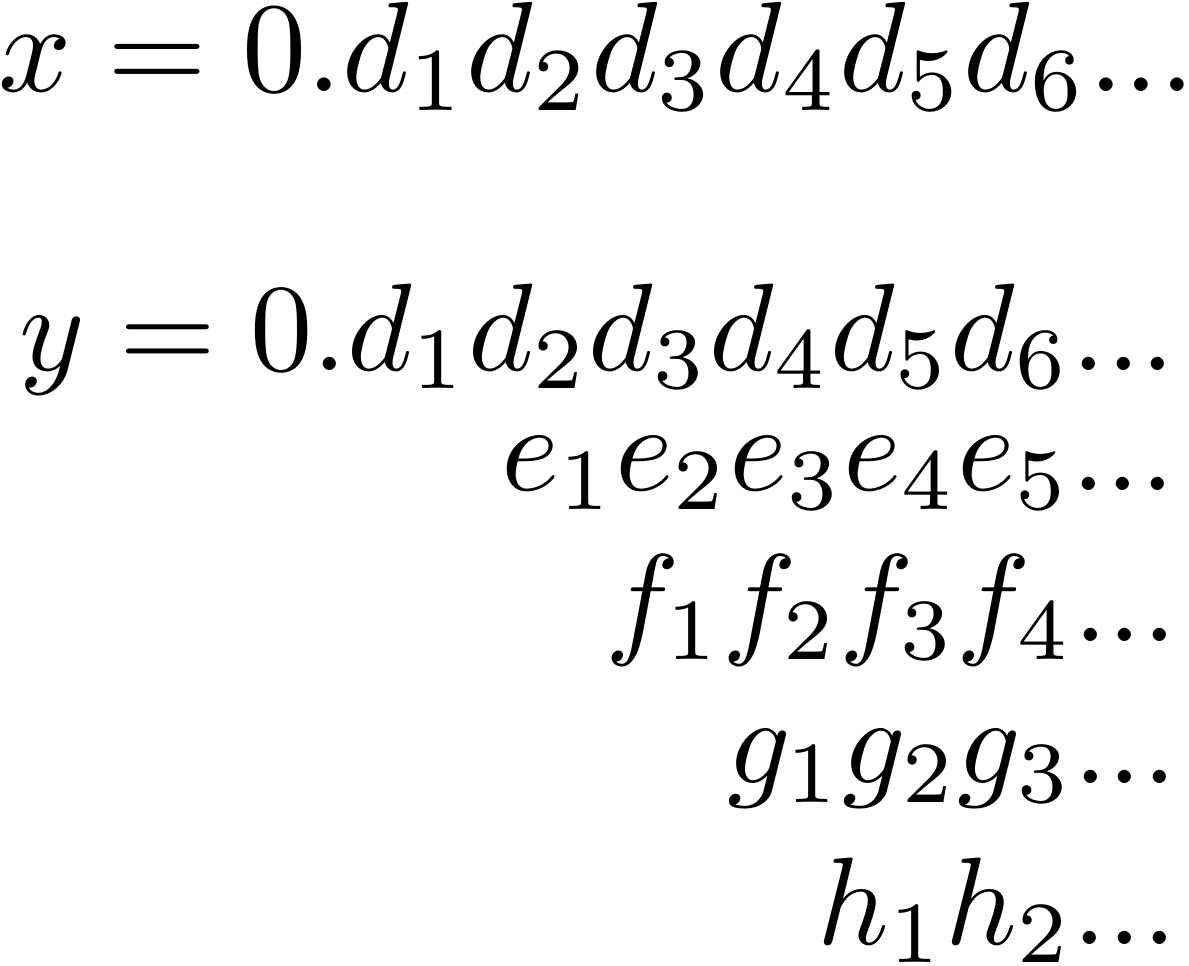}
\caption{\label{numbers}  A single real number (x, top) has infinitely many digits, which increase the amount of information available at a rate that is two possibilities for every change of scale by a factor of two. Real numbers are not good representations of systems for which the amount of information grows differently with scale (y, bottom). The number of digits as a function of scale is characterized by the complexity profile Fig. \ref{fig4}.}
\end{figure}

A single real number has infinite possibilities in all of the infinite digits of its representation. Therefore it has the ability to represent an infinite amount of information. This would seem to indicate that we could use a single real number to represent a system. For example, a number that represents the density of a liquid has infinite information, but we know from phase transitions that this single number isn't enough. Why doesn't this work? The problem is that the way the information is organized in scale in the real number does not correspond to the way it does in the system. A real number can represent the position of a point along one dimension. Let's say we start by knowing where the object is to a resolution of $1$ unit of length. Increasing the resolution by a factor of two means we can distinguish which of the two possible segments that are $1/2$ units of length it is in. Communicating this information requires a binary variable. For each $2$ fold increase in resolution we have 2 additional possibilities to specify. The number of bits is the logarithm (base 2) of the scale of resolution. However, for a liquid at its critical point the number of bits increases differently with increasing resolution. As resolution increases we have to describe the fluctuations of density. The growth in the number of bits is more rapid than one bit per factor of two in resolution (see Fig. \ref{numbers}). 

A sufficient representation, therefore, is one that has a set of possible states corresponding to the set of {\it distinguishable} states of the system at each level of resolution, down to the level we need to describe the properties we care about---the relevant parameters---and no further.
When considering interventions that affect the large scale properties of the system, rather than accumulating details about the system, we should start with the largest scale pattern of behavior and add additional information only as needed. According to the complexity profile, each piece of information about a system has a size---the largest scale at which we can begin to detect that piece of information. 

Quite generally, therefore, we are concerned with describing systems by information as a function of scale (Fig. \ref{fig4}) \cite{dcs}. Intuitively, as we increase the scale, less and less information is needed. Where there are plateaus in the information as a function of scale we have clear cases of separation of scales. At second order phase transitions we have a power law behavior as a function of scale which extends to the size of the system. There is no separation of scales.

There are ways that the use of renormalization group in physics differs from what we are interested in for complex systems and its formalization by the complexity profile. Renormalization group is about the very largest ``thermodynamic'' scale, i.e. the infinite size limit. However, in considering the behavior of complex systems we are interested in the scale of the system at which the behavior of interest occurs, i.e. human health or social economic activity. The infinite size limit is not always a useful way of thinking about these systems. Renormalization group is also described in terms of the energy of a system. For our purposes, the property we want to identify is the set of distinguishable states of the system that characterize its dynamic behavior, at a particular scale. 

Still, the methodology of renormalization group offers us more than just the concepts. We gain formal guidance about how to construct models based upon the aggregation of components to identify the relevant variables. As we increase the scale we see fewer details. Small distinctions disappear and only larger distinctions that involve many parts of the system together remain. How properties of parts aggregate determines what is observed, i.e. what is important. By studying the way that properties aggregate we can identify the important larger scale system properties. Aggregation is determined by how the parts depend on each other. 

\begin{figure}[tb!]
\centering \includegraphics[width=3.5 in]{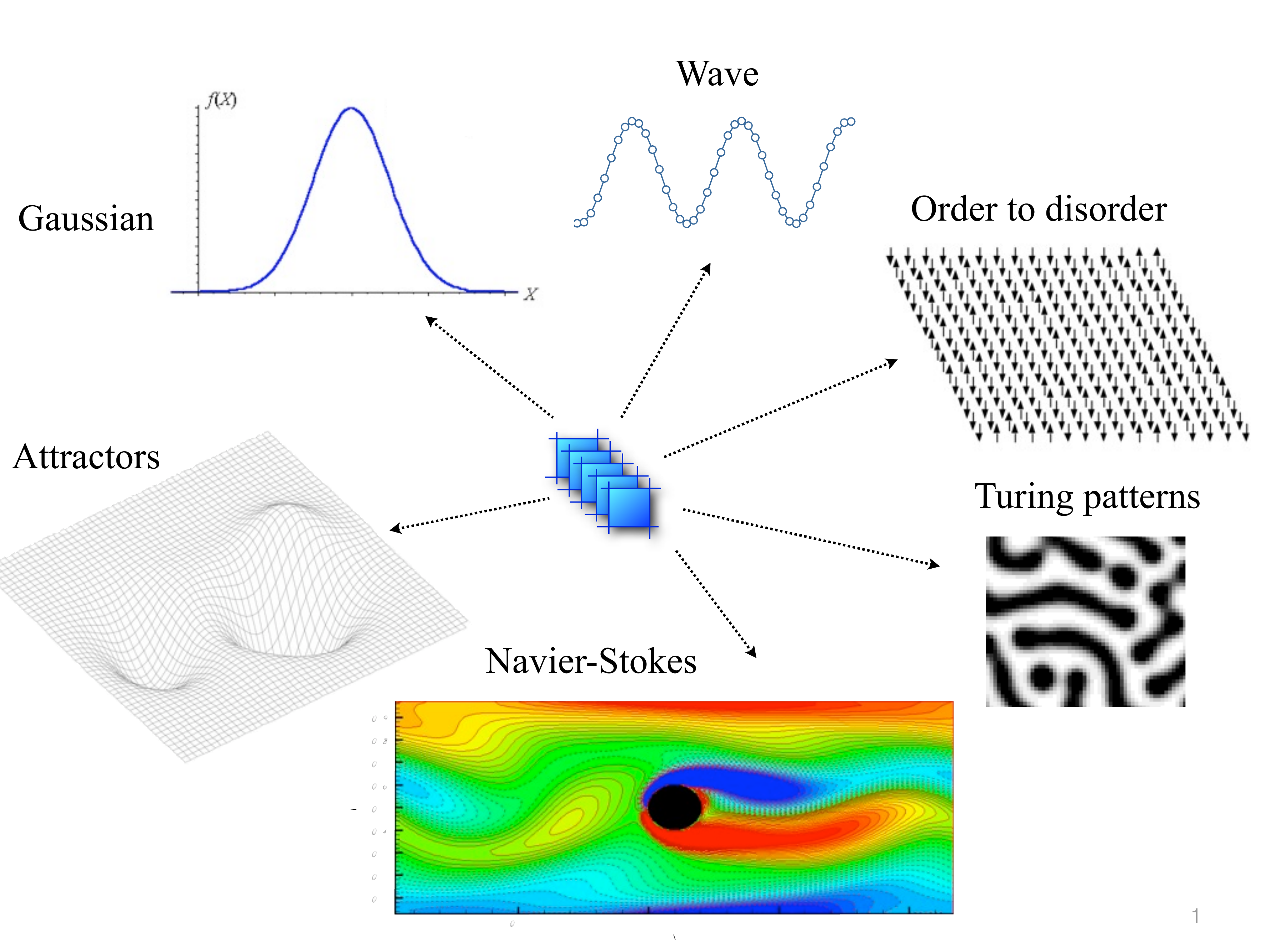}
\caption{\label{universality} When we focus on the largest scale, system behaviors map onto simplified models, each of which applies to a large set of possible systems with widely different microscopic details. Examples shown in this figure: the Gaussian distribution, wave motion, order to disorder transitions (the subject of phase transitions discussed in the text), Turing patterns, fluid flow described by Navier-Stokes equations, attractor dynamics. The existence of only a few models that capture the behavior of a wide range of systems underlies the idea of universality---systems are members of universality classes of behavior.}
\end{figure}

The simplest cases are when parts are completely independent or dependent; in the former case, aggregation is what we know from statistics that gives rise to the average and the random deviations described by the normal (Gaussian) distribution, and in the latter a single coherent behavior arises. When there are other kinds of dependencies, as explained in the next section, different behaviors occur. These include behaviors such as dynamical oscillations and spatial patterns (Figure. \ref{universality}).  

\section{universality, normal distributions and beyond}

When we go to the largest scale behaviors of a system, we simplify the mathematical description of the system because there are fewer distinguishable states, so that there are only a very limited set of possible behaviors that can happen. This also means that systems that look different on a microscopic scale may not look different at the macroscopic scale, because their mathematical descriptions become the same.

One example is that the transition when boiling a liquid to a gas has the same properties as the one that occurs when a heating a magnet up to the point where it becomes non-magnetic (ferromagnet to paramagnetic transition). 
This is neither coincidence nor mere analogy, but a direct mathematical relationship. Magnets have local magnetizations that fluctuate at a magnetic critical point just like the density at the water-vapor critical point. Just as molecules of water undergo a transition from order to disorder as the temperature increases, so do the little magnets undergo an order-to-disorder transition. The local magnetizations interact with each other just like the density fluctuations in water at the critical point. The result is that these two seemingly different types of systems map mathematically onto each other. 

The mapping of water to vapor transitions onto magnetic transitions illustrates how one type of behavior can describe many possible systems. As renormalization group was more widely applied, many instances were found of systems that have the same behavior even though they differ in detail, a concept that became referred to as {\it universality}. Still, while many systems have the same behavior, there are systems that have distinct behaviors. Together this means that systems fall into classes of behaviors, leading to the term `{\it universality class}.'  Power laws often arise in the context of behaviors that exist across scales, and the value of the exponent became used as a signature of the universality class \cite{rg,a,exponents,kardarbook}.

In a sense, the idea that many systems can be described by the same large scale behavior is used in traditional theory. Scientists use the normal distribution for many different biological and social systems. Any system having sufficiently independent components satisfies the axioms of the central limit theorem, and can be mathematically described by the same aggregate behavior, i.e. the normal distribution. When there are dependencies, the normal distribution no longer applies, but there are other behaviors that are characteristic of other kinds of dependencies. To study those behaviors, we have to determine the way kinds of dependencies give rise to kinds of large scale behavior. 

There are even more basic ways the common mathematical description of systems is used, i.e. point particle motion is used to describe the motion of many distinct objects, and wave equations are used to describe everything from music strings to water waves to light. 
Even though the specific systems are very different, the dependencies that give rise to their behaviors, and the behaviors themselves, are related mathematically. 

We can consider universality even more generally for complex systems. To generalize the ideas of universality arising from renormalization group, we do not consider the macroscopic limit of infinite system size and power law exponent. Instead, the states of our representation must correspond to the states of the system at the scale of observation. Moreover, instead of using those states to describe the free energy as is done in physics of phase transitions, we use them to describe dynamics, and the response of the system to external forces. Underlying any equation is an assumption that the variables that are being used are a valid (sufficient) representation of the system. The relevant parameters are those that are important at a particular scale of observation, determined by distinguishability to an observer. The mathematical representation of one system at a particular scale may also correspond to the behavior of other systems. The systems may have quite different underlying components.  This is a general concept of universality (Fig. \ref{universality}).

What are the cases where the thermodynamic limit does not serve to expose universality? An important example is pattern formation, as found in Turing patterns \cite{turing}, spatial arrays of roughly, but not exactly, periodic spots or stripes. These patterns can arise in many ways, for example from the reaction of diffusing chemical species \cite{epstein}. At large enough scales, these patterns look only gray. This is a case where the system behavior occurs over a limited range of scales, and the degrees of freedom disappear at large enough scale. Still, we can map these descriptive variables from system to system. These patterns represent universal classes of behavior for systems that have structure at a particular scale. What is important is that the patterns arise from a wide range of possible microscopic behaviors, but they are not sensitive to the details because the scale of the pattern is much larger and therefore separate. While there are parameters that are needed to describe the variety of patterns that are formed, there are far fewer of those than in the microscopic material details. Microscopic changes of the substances that comprise them only change the pattern to the extent that they change the relevant parameters, and only in ways to be understood by how they affect those relevant parameters. 

Including Turing patterns in the concept of universality is a natural and important generalization of the concepts of universality in statistical physics. The changes of these patterns with relevant parameters, and their response to environmental conditions, can be studied and related to phenomenological observations. There is predictive capability for interactions at those scales of observation. These issues are not trivial, as they present challenges for mathematical and observational studies. The adoption of Turing's ideas in biology for the description of patterns on animal skins and morphogenesis more generally has been controversial \cite{anti,pro} precisely because the pattern dynamics does not capture microscopic mechanisms.
This controversy misses the key point about universality. Universality should be intuitive in this context as the details of molecular processes are not necessary to characterize the variation between individual animals with similar patterns, or between species, or the dynamics of a pattern as it forms, and do not affect roles of these patterns in social and ecological interactions. Once the description of the behavior of the system is identified in terms of relevant variables, any genetic or environmental influence on those patterns must affect those relevant variables. The behavior can be studied in terms of the relevant variables as well as the underlying mechanisms that determine the origins and values of the relevant variables. This is similar to the ability to describe planetary motion without describing details of individual planet structure. 

The study of universality enables us to identify classes of systems whose behaviors can be described the same way and can be captured by a common mathematical model. This is the principle of universality that is formalized by the analysis of renormalization group and generalized by the application of multiscale information theory to the scientific study of complex systems. A good way to think about this is that the mathematical model describes one member of the class. 

\section{Dynamic sensitivity and chaos}

When we consider the scales of behavior of a system, we also have to address amplification and dissipation, which enable smaller scale changes to impact on larger scales (amplification), or larger scale behavior to become confined to smaller scales (dissipation) over time. In such cases, the amount of time over which a system is being studied determines the level of detail at one time that is needed to describe the large scale behavior of the system at a different time. The amplification or dissipation rate can be related to the extent of additional information that is needed to describe the system at a particular scale as a function of time scale of observation. 

Deterministic chaos (e.g. Section 1.1 in \cite{dcs}), popularized as the butterfly effect in meteorological systems, in which small initial differences diverge exponentially over time, is a more specific way that scales mix over time \cite{metzlerinfo}. The divergence of trajectories is often discussed in terms of the problem of predictability because there is a limit to the accuracy of observation of a system. Limited information about initial conditions implies that later states of the system are not determined by observations at the initial time even though the system is deterministic. Chaotic systems have different degrees of predictability. Planetary orbits are chaotic \cite{sw}, despite high levels of predictability. What is important for predictability is the rate of divergence, which is measured by the Luyaponov exponent, or the time scale at which divergence at the largest scales occurs. For planetary orbits this time scale is millions of years. 

From the point of view of multiscale information theory, amplification, dissipation and chaos are aspects of the mathematical models necessary for real systems. While some consider the purpose of modeling to be prediction, the purpose of a mathematical model should be understood to be fidelity of the model to the system. Mathematical models of chaotic systems can be constructed just as other models can be. The difficulty that is encountered with predictability is not the problem of the mathematical representation. The limits to prediction are inherent in the system behavior. The objective is to obtain the best understanding that is possible---recognizing inherent uncertainty is essential. The ability to obtain the best possible understanding should not be confused with having complete predictability. (More subtly, the analytic representation of deterministic chaos assumes a set of real number parameters describing the system. However, as described in Fig. \ref{fig4} the complexity profile generally implies that there are more degrees of freedom at finer scales, and therefore the standard model of chaotic behavior does not extend to finer scale descriptions of a system. Well above the quantum limit, to which chaos does not apply, other models are likely to be needed.)

Applying the complexity profile to biological and social systems presents challenges associated with amplification (unrelated to chaos) that are important to recognize. Consider micro to macro connections in biology. A single genetic mutation in the $\beta$-globin gene, substituting valine for the glutamic acid at position 6 (or 7 including the start codon), 
causes abnormal hemoglobin molecules. One such mutation in an individual causes sickle cell trait, which provides resistance to malaria, while two such mutations lead to sickle cell disease.  Consider micro to macro connections in society. A single individual's ideas and influence can have dramatic effects, e.g. Steve Jobs' role in the development of modern devices from the personal computer to the iPhone, changing the way hundreds of millions of people work and play. These large scale differences in the physiology of an organism or a society over time might seem to differentiate biological and social systems from physical ones, though similar sensitivities to small scale events affect physical systems. Significantly, the extent to which fine scale information can affect large scale behavior and the conditions under which it occurs are limited. Understanding these conditions and the details that can be influential is essential to scientific understanding. 

The large scale impacts of mutations and ideas arise from the possibility of amplification by informational replication over time that enables them to achieve large scale. Biological mutations can have large impacts because of the replication of DNA throughout the cells of the body, and across populations, and through the subsequent transcription of their information to many proteins that function in particular ways. Social systems have sensitivity to specific ideas as they are transmitted to others, embodied in machinery and organizational processes that mass produce and distribute them widely. These are in many ways similar to the butterfly effect attributed to meteorological systems. The conditions of amplification require an available energy source in heated oceans and its ability to drive the creation of the highly redundant large scale motion of a hurricane. 

These processes are not counter to the framing of information importance. However, they do make its application more challenging as we need to understand the way that information is replicated over time. That there is sensitivity to micro information does not mean that all micro information can or will end up as large scale system behaviors. Not all molecular changes, or genetic mutations, have a large scale impact. Ideas and individuals that can change society as a whole are rare compared to the many that are present. Whether and to what extent a large scale behavior is sensitive over time to the impact of small scale events---which can have increasing impact as they are replicated---is part of the analysis of the scale of information in the behavior of a complex system. In every case, identifying the large scale information is essential to the analysis and our eventual understanding of the system. Focusing on the large scale behaviors still must be done, as mapping all of the many fine scale details at the atomic scale cannot ultimately be effective. Moreover, it is understanding which pieces of microscopic information are the ones that actually influence the large scale behavior that provides insight into the behavior of the system. 

\section{Complex systems: Examples}

The objective of the multiscale information approach is to represent correctly the largest scale behaviors of the system, leaving out microscopic details that are not essential to answer particular questions and for which reliable representation is often impractical. Representing the largest scale behaviors means identifying the set of states that are possible and their dynamics and influence by external forces. The complexity profile provides a measure of the number of states that are needed, however specific methods may identify the correct model with or without its assistance. Because of universality, classes of macroscopic behavior previously found in the description of physical systems may also be applicable to biological and social systems. Universality class analysis provides shortcuts to the characterization of either specific systems or problem domains. The examples in this section largely make use of universality classes and insights that have been gained previously. Extending these approaches to a more complete set of social challenges and health interventions will require much greater efforts. Examples provided here are only first steps in the process. 

A multiscale strategy that has been widely adopted in the study of complex biological and social systems is that of identifying power law scaling behavior in empirical data. This has been done in such wide ranging topics as economic time series, DNA sequence correlations, heart beat intervals, network connectivity, physiological and urban properties \cite{mandelbrot,StanleyDNA,StanleyHeart,MantegnaNature,barabasi,west,westcities}. Because power laws are scale invariant, such multiscale behaviors can be identified, and models developed for them, without reference to much of the framework of renormalization methods. 

The breakdown of spatial homogeneity, i.e. smoothness and averaging as mathematical approximations, in the study of phase transitions is an insight that can be expected wherever statistical averaging is being used to describe the behavior of systems. We can consider averaging not as a failed approach, but as an approximation, which may or may not be valid, and may be useful even when it isn't strictly valid. This approximation is often called a ``mean field" approximation. The reason is that the local behavior of a system is a response to the forces that act upon it, which is called the local ``field.'' If we have a model that uses the average of the local fields across the material instead of the actual field everywhere, then not only the field but also the local behavior will be the same in the model. This becomes an approximation to the large scale behaviors of the system. Indeed, the breakdown of this approximation is a useful demarkation between systems that are describable within traditional statistical tools and those that require complex systems methods, i.e., where dependencies in the system give rise to multiscale behavior. Thus we can expect that in biological and social systems there will be many cases where the breakdown of the ``mean field" approximation provides new insights.

The first example we give is to the mathematical analysis of evolutionary theory in biology with implications for biodiversity, altruism, speciation and longevity. 
More specific examples follow of applications to social systems, i.e., ethnic violence, global food prices, panic and organizational effectiveness. 

\subsection{Evolutionary dynamics}

Statistics was developed in conjunction with the NeoDarwinian view of evolution by Fisher \cite{fisher}. In the 1920s, statistics was a powerful new approach to describing how parental genomes combine to become those of offspring (Mendelian inheritance). In the approach he used, each offspring is treated as one instance of all possible combinations of parents. The mathematics he developed continues to play a central role in the analysis of heredity and evolution of traits, i.e. population biology.   
However, the statistical methods he used are an approximation whose significance is still not widely understood. 

This statistical analysis in population biology can be mapped directly onto the mathematical issues discovered in the study of phase transitions in physics. The analysis of evolutionary dynamics differs only in that there is a dynamical equation instead of the equilibrium description of materials. In each case the mathematics begins from a description of the probability of states of the system, and this is related to the states of the components. 

Just like regions of a material can be in different states, e.g., \{liquid, gas\}, organisms in a population can have different possible genomes, i.e. sets of alleles. The entire population is a combination of individual organisms, which are combinations of genes, each with its own state (one of several alleles). The assumption that Fisher made is that one gene can be assigned an allele independently of the specific state of the other alleles of that organism and of other genomes. A subtle point is that it isn't independent of the {\it set of states} of the genome of other genes of the same organism and other organisms (i.e. which alleles are present in the population, which is essentially the same as the {\it probability} of encountering them in a random sample). The state of one gene is independent of the specific state of other genes or genomes. Mathematically, the probability of the genomes of all the organisms can be written as the product of the probability of the states of individual organisms, or even individual genes. 
This is a general way to make the separation of scales ``mean field" approximation,
i.e. $P(\{s_i\}) = \Pi_i p(s_i)$, where $P(\{s_i\})$ is the probability over the state of the system as a whole (the material, or the genomes of all organisms), and $p(s_i)$ is the probability over the individual component (a small area of a material, or an individual gene) of taking a particular state. This makes apparent that dependencies are approximated: the specific state of one of the component does not affect the state of another, only the probability of the states of the individual components affects the other components.   
This distinction is a subtle but essential mathematical one.
 
To make the discussion easier to follow, we will use the ``rowers' analogy" used by Dawkins in his book {\it The Selfish Gene} \cite{dawkins} to explain both the statistical approximation and its breakdown. 

In the rowers analogy, we think about races between teams of rowers in boats. A rower is analogous to a gene, and a boat is analogous to an organism. There is a Òrower pool,Ó with rowers that are placed into boats, with all boats having the same number of rowers. The boats run heats against each other and the winners are placed back into the rower pool to compete again. To make up for the rowers that lost so that we always have the same number of them, the rowers Òreplicate,Ó i.e., the the number of successful rowers are increased in number while retaining the same qualities. 

An example Dawkins describes is a competition between English speaking and German speaking rowers. The languages affect the race because one-language boats have an advantage---the rowers can understand each other---and win. What will happen over time to the rower pool? If there are more English speaking rowers, there is a higher probability that a boat will have all English rowers. Moreover, German speaking rowers will tend to have English speaking partners. This means that English speaking rowers will win the race more often than German speaking rowers. Over time, the number of English speaking rowers will grow and the number of German speaking rowers will shrink. Eventually there will be an all English speaking rower pool. Alternatively, if we start out with a rower pool that has more German speaking rowers, over time the number of German speaking rowers will grow, and we will end up with an all German speaking rower pool. In either case, we can think about this as a competition between the rowers, with one type of rower winning over the other type over time.

Dawkins' argument seems reasonable, but a hidden assumption 
has surprisingly far reaching consequences. The assumption is hidden in how rowers are placed into and taken out of the rower pool. He assumed that this would be done at random. What happens if we don't do this? Consider, for example, what would happen if we have a rower pool as a line of rowers. We take rowers out of the front of the line and put them into the boats, then place the rowers that win back into the rower pool at the back of the line. The dynamics would be very different.

At a certain place along the line of rowers, rowers will tend to become the same type, English or German speakers. However, the type that is found may be different in one part of the rower pool line than another. 
Patches that arise lead to a multiscale structure of the population! The existence of patches makes the process quite different from the mixed rower pool. Moreover, in describing the system it is not enough to recognize that the are patches, but also that the patches change over time due to the movement of their boundaries. The dynamical description of the system includes the multiscale dynamics of those patches. 

One way that process is different is that both English and German speakers persist for a very long time. It might happen that one or the other of them will disappear eventually, but it will take many more generations for this to happen if we have a line as a rower pool than if we mix them up every time. Interestingly, this might also be the reason that there are English and German speaking people. If everyone was mixed frequently in the world, then it would make sense to have only one language, but if people who speak German live in one part of the world, and people who speak English live in another part of the world, it is possible to have multiple languages, where some areas are English speaking and some are German speaking. It makes sense that today when people move more than in the past, there is more of a tendency to speak a single language than when people didn't move around as much.

The difference between taking the rowers at random and not at random can be understood in terms of the probability distribution of the rowers in the boats. Taking them at random means that the likelihood of a particular rower is given by the probability of the rower type in the rower pool, independent of other rower types in that boat or other boats. This is exactly the mean field approximation. When we take them from a particular part of the rower line, the probability of one rower type is not independent of the type of the previous and subsequent rower that are chosen along the line because they all come from the same part of the rower pool that is much more likely to have the same rower type than the rower pool as a whole.  

The biological analogy of the linear rower pool is the mating of organisms that are nearby to each other geographically (or alternatively mating according to traits that lead some organisms to be more likely to mate with others of similar types, assortative mating). It is enough for animals or plants to reproduce near to the place where they were born to change dramatically the conclusions of Neo-Darwinian theory. This change is surprising to many who are not familiar with the difference between mean-field and non-mean-field behaviors. 

A first result of realizing that biological populations are not described by the mean field approximation, is that populations can be much more diverse than is predicted by conventional population biology \cite{baryam,sayama}. Biodiversity was a central motivation of Fisher's work, as prior theories could not describe the high level of biodiversity found in nature \cite{fisher}. Going beyond traditional statistics improves on his results, as even greater diversity arises from the breakdown of the approximations he introduced. In a well mixed population the diversity disappears exponentially fast. It persists much longer when it is not well mixed. Interestingly, many of the experiments that test population biology are done in a laboratory where populations are mixed. So the assumptions of the theory and the experimental conditions match. Laboratory populations are known to be very homogeneous in genotype and natural populations, ``wildtype," are much more diverse, consistent with the expectation that they are not well mixed  \cite{wild1,wild2,wild3,wild4}.

A second result has to do with a specific form of diversity, speciation, the separation of one species into two or more species over time. The process by which this takes place has been a subject of much controversy. This can be understood because if we start by considering the species to be mixed at every generation of mating, then how does it stop being mixed? If on the other hand we include a non-mean field description of a species, then speciation results from progressive separation of types that are more likely to reproduce with those of the same type \cite{doebeli}. The spatial population version of this idea has been shown to describe natural biodiversity very well \cite{aguiar} by considering spontaneous spatial pattern formation of patches of different types that separate genetically to become species in a high dimensional type of Turing pattern.  

A third result has to do with altruism and selfishness, and the primacy of competition over cooperation, in evolution. One of the key results that was gained from Fisher's statistical methods is the idea that is popularized by Dawkins, the ``selfish gene.'' According to this view an individual gene has traits that only benefit the likelihood of its own reproductive success, and not, for example, that of the entire genome, family, or species. The idea that genes are self-interested is proven using the statistical assumptions of Fisher. An important question is whether this conclusion is valid when those assumptions do not hold. The answer is that they do not \cite{rauch,werfel,stacey}. The self-generated patchiness of the population creates new dynamical processes that are not described by selfishness of genes. Thus, the breakdown of the mean field approximation is important for understanding why competition between genes is not enough to describe evolutionary processes. This issue has taken the form of a raging controversy about altruism. If the ``selfish gene'' view holds, then the only way that altruism among organisms can arise is if they have the same alleles (one organism then helps another to increase the number of its own alleles in the next generation) and the extent of that altruism is limited by the extent of the sharing of alleles. Mathematically, this idea is embodied in a mechanism called ``kin selection'' and there is a controversy today about whether kin selection is enough to describe evolution, or whether the role of association in groups is needed, a theory termed ``group selection'' \cite{wildwest,wade,nowak,abbott}. Often the controversy is discussed only in the mean-field approximation, in which case the two theories are mathematically equivalent and the argument becomes one of concept rather than mathematics. When we go beyond the mean-field approximation we find that kin selection is not sufficient.

We can see how this works from the rower pool idea. Consider a combination of altruistic and selfish rowers. If there are patches of altruistic and patches of selfish individuals in a linear rower pool, the selfish individuals do worse than the altruistic ones. This is because the altruists are generally near altruists, and the selfish individuals are near selfish individuals. Selfish individuals are unable to take advantage of the altruists as they are not next to them in the rower pool and thus don't end up in the same boats. This would not be the case if the rowers are selected at random. Still, there must be boundaries where selfish individuals are near altruists, and we need to understand what happens at the boundaries. Thus the evolutionary dynamics is controlled in important ways by the properties of boundaries, i.e. boundaries are ``relevant," which is not the case at all for the NeoDarwinian approach that only focuses on the properties of alleles. Results that depend on how patches and boundaries work are not described by kin selection and lead to a different understanding of altruistic behavior. 

A fifth result has to do with the implications in the real world of increasing rates of travel using airplanes and other means of transportation globally \cite{diseases}. We saw that mixing leads to mean field type of behavior that is very different from the case of spatial models. Increasing transportation brings us closer to the well mixed case. Biodiversity should decrease. We see this effect in that invasive species are eliminating local variation of species. Another effect is a change in the pathogens responsible for infectious diseases. When a pathogen can only spread locally, the most aggressive strains go extinct as they kill off the local hosts in their local patch. As transportation increases patches become irrelevant and the successful strains are those that are more aggressive and deadly. The practical implications are present in the increasing prevalence of deadly pandemics. A theoretical analysis of the increase of long range transportation implies that the increasing severity will lead to global extinction, where local extinctions were present before. Implications for pandemic response are important for Ebola and other severe diseases. 

A sixth result is in the study of evolution of longevity, with implications for the ability to extend lifespans \cite{longevity}. In this case the mean field approximation suggests that evolution cannot select for specific lifespans due to the benefit to reproduction of living longer. However, local reproduction links organisms to their descendants so that selection of lifespans is determined by ecological conditions affecting multiple generations such as the local availability of resources. Because lifespans are selected by ecological conditions, interventions in those mechanisms may result in extended lifespans.  

Thus we see that there are many problems with the Neo-Darwinian view of evolution. It is a useful and powerful approximate way to think about evolution. However, finding out that it is not always correct includes the realization that biodiversity, speciation, altruism, disease, longevity and other important domains of biology are not fully described by it. 

\subsection{Multiscale biodiversity}

A direct application of the complexity profile has been made to analysis of biodiversity. The evaluation of complexity and scale in this context has direct relevance to our understanding of the vulnerability of species when there are losses of diversity.  

The loss of biodiversity is a current challenge for global conservation efforts due to direct exploitation, reduction in natural habitats, and other effects such as global transportation of invasive species. Global efforts to conserve species have been strongly influenced by the heterogeneous distribution of species diversity across the Earth. This is manifest in conservation efforts focused on diversity hotspots \cite{prendergast,myers,gaston}. The conservation
of genetic diversity \emph{within} an individual species \cite{wilson,faith} is an important factor in its survival in the face of environmental changes and disease \cite{amos,frankham}. A multiscale characterization of biodiversity has shown that diversity within species is also unevenly distributed. Genetic distinctiveness has a scale-free power-law distribution implying that a disproportionate fraction of the diversity is concentrated in small sub-populations. Diversity has its own internal dynamics, which are distinct from possible outside influences such as habitat change and species interactions. Increases happen only gradually, but large decreases may occur without an extrinsic perturbation due to death of rare types. 

To analyze the biodiversity it is helpful to consider not just the variety of genomes present in a species, but the number of repetitions of a particular type. This is the converse of the complexity profile, i.e. if the complexity profile is the variety as a function of scale, the multiplicity of genomes is the scale (redundancy) as a function of the variety. Multiplicity is a measure of the population structure. Multiplicity can also be defined using the number of members of a species, i.e. as a measure of the scale of the range of genotypes present in a species. Biodiversity is then evaluated either as a measure of the distribution of the multiplicity across species \cite{allen}, or within a species across types.  

An example of the application of multiplicity in the evaluation of the importance of biodiversity in the robustness of biodiversity to catastrophes. It has been suggested that extinction of 95\% of species would leave 80\% of the tree of life (total diversity) retained, and that as a result ecological planning to preserve diversity is not constructive\cite{nee}. These results were obtained because random losses, even when high, are unlikely to remove all individuals belonging to a deep branch of the tree even when it forms a small proportion of the population, thus preserving most of the diversity. In contrast, the analysis of multiplicity suggests that conservation planning is important and can enable substantial improvement of diversity preservation. The small immediate loss is followed by a much greater loss over time due to the vulnerability of small residual populations to extinction. The vulnerability of residual small groups of related organisms can be seen from the multiplicity of the population---both scale and diversity matter!  
The loss of a large fraction of a group of closely related species (or of closely related organisms) leaves the remainder of the group highly vulnerable to extinction. Thus, while the survival of at least one organism of a group is probable, the viability of that type is compromised and subsequent extinction is likely. This result arises because of the non-robustness of small scale diversity in contrast to the robustness of larger scale diversity. It shows that ensuring the reproduction of rare types by conservation planning \cite{amos,faith,crozier,moritz} during or even just after an extinction episode can dramatically improve diversity retention. 

\subsection{Ethnic violence}

Complex socio-economic systems might seem to be difficult to study using universality and renormalization group / multiscale information methods. However, several examples exist that show that it is possible to apply these methods. The first example helps clarify the origins of ethnic violence and its prevention \cite{lim07,rutherford11}. 

Ethnic violence is often described in terms of historical, economic, political, leadership and various other social aspects of the conditions of those who engage in violent conflict. Ethnic violence is typically a collective behavior involving multiple individuals that decide upon and engage in violence rather than due to an individual decision maker such as a national political leader. This is a signature that an analysis of the relevant large scale (collective) parameters is useful. How might we apply the arguments of universality in this context? The objective is to determine a few measures of a social system that are relevant to whether or not ethnic violence occurs and thus to help identify where or when it happens and how to prevent it.

This has been done by building a theoretical framework that considers solely the existence of distinct types, without explicitly treating other factors. Such a framework is consistent with reducing the analysis to only the essential descriptors of a problem, i.e. without distinction between types there cannot be ethnic conflict, and there is no other condition that is, a-priori, essential. The other factors may play a role but they are assumed to be linked to the dynamics of the types.  As discussed above in the treatment of phase transitions, the presence of spatial dimensions is also relevant and since human populations are found in an approximately two dimensional space, we include these dimensions. 

The most general model with these attributes has types that are spatially distributed and follow a dynamics that involves movement in space. There are then two different types of behavior, mixing and separation. Considering only local movement the behavior is that found in alloys of materials. Alloys can mix or separate. Separation occurs because of an energetic (atoms) or social (human) preference for being near members of the same type. Separation manifests in the existence of progressively larger groups over time. Universality means that microscopic parameters do not influence the behavior except in a multiplicative constant that controls the rate at which the progressively larger groups form \cite{bray}. The only descriptive parameter of the system is the patch size. For systems where migration can lead to larger scale movements of individuals, those movements lead to the existence of patches whose size is, for our purposes, extrinsically imposed. 

Given the existence of only a patch size as the relevant parameter, it is necessary to consider how ethnic violence may be affected by this size. There is no prescription from physical law that specifies this dependence, but intuitively it has been argued that in either of two limits violence will not be present. First, when neighborhoods are well mixed. This, in effect, reverts to the case where individuals do not choose to separate, or to conditions in which everyone knows members of the other groups personally, and in this limit we can assume that widespread violence should not be present. In the opposite limit in which patches are large, the overall condition is that individuals do not see members of the other type and therefore there should not be spontaneous violence. This suggests that violence will be present only at a particular intermediate size of patches. Having identified a single relevant parameter as a patch size that promotes violence, it remains to validate the analysis against observations of actual violence.  

\begin{figure*}[tb]
\centering \includegraphics[width=4.5in]{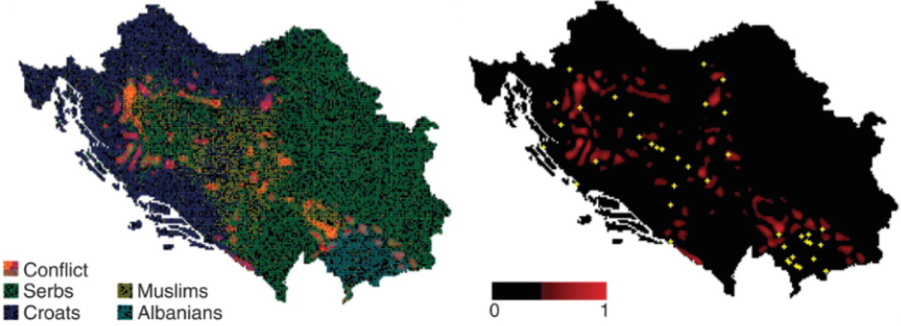}
\caption{\label{figethnic} Yugoslavian census data from 1991 were converted into a spatial representation [left] and used in an agent-based simulation to predict populations likely to be in conflict with neighboring groups [red overlay, left and right]. The prediction agrees well (90\% correlation) with the location of cities reported as sites of major fights and massacres [yellow dots, right].}
\end{figure*}

Comparison with data on ethnic violence in the former Yugoslavia (Fig. \ref{figethnic}) 
and India confirms \cite{lim07} that detection of patches of size 20-60 km provides a high spatial correlation (e.g. 90\%) with locations of ethnic violence. 
The results can be explained sociologically because separation limits inter-group friction, while integration inhibits inter-group alienation. In places where self-identifying groups separate into geographical patches of the critical size a group begins to impose its cultural norms, religious values, language differences and in-group social signaling within public spaces. These spaces may include public squares, markets, restaurants, places of worship and schools. However, when social expectations are violated by members of other groups, the resulting friction is likely to cause radicalization of some members of the population.  
For patches larger than the critical geographical size, individuals remain largely within their own domains and de facto local sovereignty exists. If patches are smaller than the critical size, ethnic groups cannot impose their own norms and expectations about behavior in public spaces, allowing for the peaceful coexistence of the multiple ethnic groups that are present. This discussion further suggests that natural and political boundaries can increase autonomy to allow for separation that can prevent violence in areas where it would otherwise occur. An analysis of Switzerland shows that this is consistent with both historical events leading to the formation of its Canton based federal governance structure and with current census geographical data about linguistic and religious groups, including modern day peace and violence \cite{rutherford11}. 

The agreement between theory and observations for Yugoslavia, India and Switzerland provides an important confirmation of the approach we are using to identify relevant parameters in highly complex social systems to enable theories that can identify both the locations of and mechanisms of ethnic violence and its prevention. The research further demonstrated that well-placed intrastate political borders and well-situated topographic boundaries that promote local autonomy (as in Switzerland) can reduce the propensity to violence, providing a method for a more limited intervention than full integration or separation to promote peaceful coexistence when it is necessary \cite{rutherford11}, an approach that may be considered in many violence prone areas of the world. 

Other social and economic forces can be considered in the context of the model. First, the identification of geographic patch size as a relevant parameter allows not just for violence to be associated with patch size but many other social properties. Various economic and social conditions can be expected to be linked to the patch size as well. Correlations of these properties with violence as well as with each other, can expose those dependencies if the patch size is used as an independent variable.

Second, just as migration and political barriers were treated extrinsically, other external forces can affect the behavior of the system. Such larger scale forces may suppress or promote violence. A relevant example is the role of dictators, such as Tito in Yugoslavia, whose regime inhibited the violence that arose afterwards. Another external force would be collective decisions to impose political boundaries, as has been done in Switzerland, or decisions to enforce mixing in public housing in Singapore. These are not only extrinsic forces, they, in effect, make use of the understanding that is found in the theory to act on the system itself. As in physics, the understanding of system behavior enables interventions to modify those behaviors. 

We note that the identification of a geographic patch size of a population as a parameter relevant to ethnic violence is a scientific hypothesis whose validation by empirical data is a confirmation of the validity of that hypothesis. However, the generation of that hypothesis by the multiscale information (renormalization group) approach is also a  confirmation of that methodology as a way of generating ``hypotheses'' for a wide range of complex systems. 
This approach is important precisely because identifying the right (low dimensional) hypothesis for testing in a complex system is itself a task whose successful performance would be expected to be highly improbable, i.e. there are a large number of potential hypotheses and testing each one of them would limit the validity of statistical tests due to the problem of spurious correlations. In this case, as in other applications of multiscale analysis, the identification of a unique hypothesis provides a strong theoretical basis for complex systems understanding. 

\subsection{Dynamics of market prices}

For a first economic example of the multiscale information approach, we review a recent analysis of bubbles and crashes in the commodity markets \cite{food_prices,Feb_update} that matched actual price behavior precisely. 

\begin{figure*}[tb!]
\centering \includegraphics[width=4.0in]{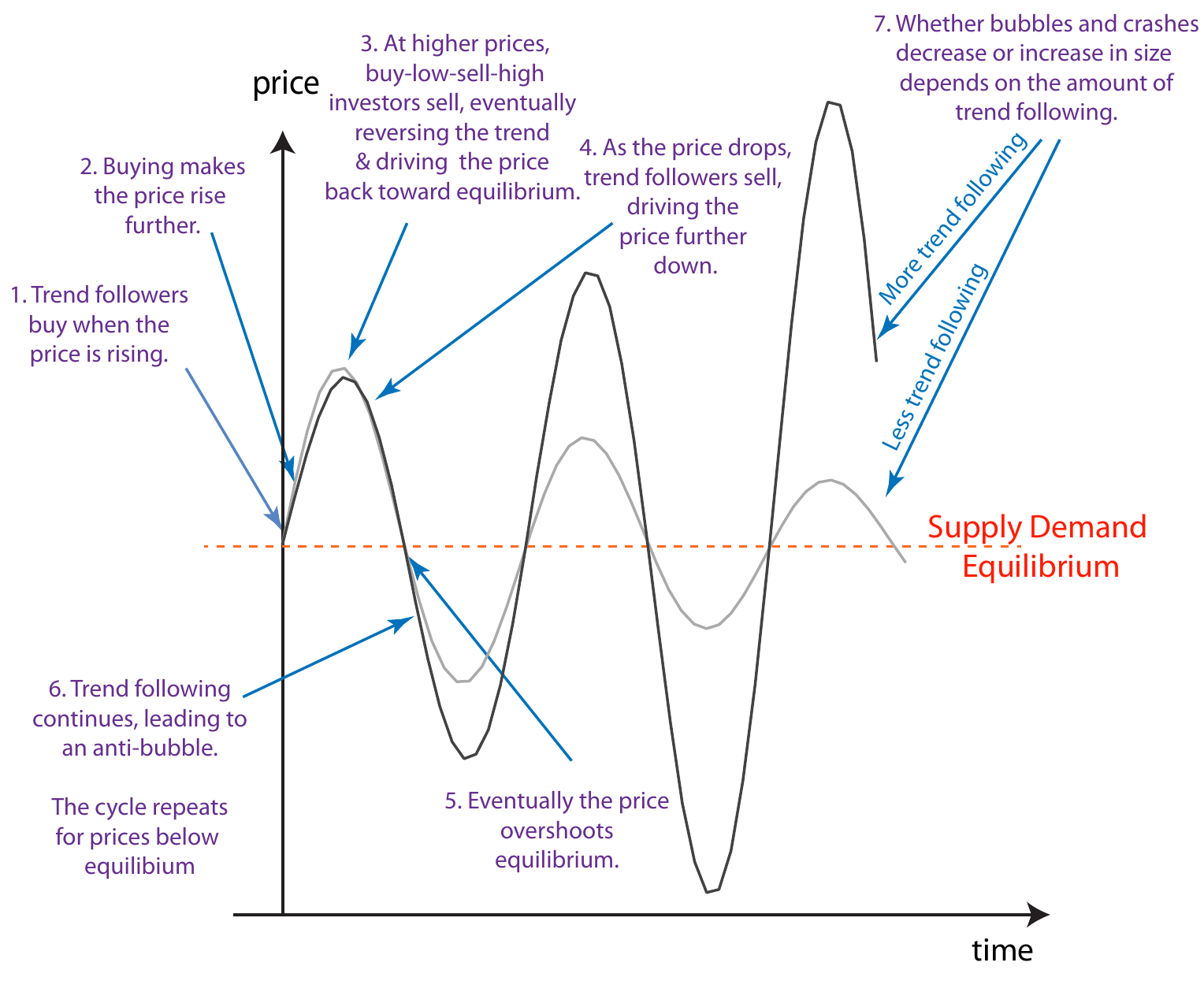}
\caption{\label{fig1} Bubbles and crashes in market prices result from bandwagon effects when price changes themselves motivate traders, leading to mutual influence between traders. Thus, buying can cause more buying and selling can cause more selling. More precisely, bubbles arise as an interplay between two different kinds of investors, trend-following speculators who buy when prices are increasing and sell when they decrease, and fundamental investors who buy-low-and-sell-high based upon supply and demand information.}
\end{figure*}

The traditional theory of markets assumes that people decide on investments independently and rationally, and therefore predicts a supply and demand equilibrium. Interestingly, in a complex systems analysis of market bubbles, it is not so much the assumption of rationality that fails, but rather the independence. Due to trend-following in the commodity markets, actions of individuals are not fully independent; rather people make decisions that influence and are influenced by the decisions of others. These influences lead individual actions to combine into collective oscillations (Fig. \ref{fig1}). This example illustrates a case where the large scale model is of a time series, i.e. the dynamical behavior of a system. Renormalization group approaches to dynamical stochastic equations have been described in physics in the study of material growth \cite{kardar}. Our application is simpler than the physics problem since there is only a single dynamical variable and stochastic terms are not necessary in this specific case, but the framing is the same.  

To construct a dynamical model of food price changes, we write the incremental change in price as:
\begin{equation}
\begin{array}{ll}
P(t+1)-P(t) = &- k_{sd} (P(t)-P_e(t)) \\ 
&+ k_{sp} (P(t) - P(t-1))
\label{eq:model0}
\end{array}
\end{equation}
which is an expansion to first order in variables describing the system; these are the largest terms affecting the behavior. Using only the largest terms is justified by the multiscale information methodology as global food prices can be affected only by many individual components acting collectively at the largest scales. The first term can be identified with buy-low sell-high investors with a fundamental price determined by supply and demand, $P_e(t)$. The second term can be identified with trend following speculators, who buy when the price goes up and sell when the price goes down. The first term gives a dynamical version of the traditional equilibrium market, and is extended to include relevant terms that lead to intrinsic self-generated dynamical price behaviors. In a more complete model, additional terms are added for the tendency of trend followers to switch markets. When prices change dramatically, speculators move to the market with price increases, and away from one with price declines. 

Without the trend following speculators, the action of the buy-low sell-high investors leads to an exponential decay toward the equilibrium price. With the trend following speculators, the system has different behaviors depending on the extent to which they exist, i.e. the magnitude of the coefficient $k_{sp}$. If the price of a commodity happens to go up, trend followers who buy when prices go up push the price further away from equilibrium. However, the further the price is from equilibrium the more the conventional investors who ``buy low and sell high'' get involved, and their selling provides a force driving the price back towards equilibrium. Indeed, the further the price deviates from equilibrium, the more this force driving back toward equilibrium (known in economics as a ``Walrasian force") strengthens, eventually reversing the upward trend. At this point, the trend following bandwagon effect drives the price down, eventually overshooting equilibrium and restarting the cycle. This interplay results in an oscillation of prices deviating from equilibrium. Rather than calculating an unstable equilibrium price, a methodology including interdependence identifies the large scale pattern of the system and accurately maps it onto the bubble and crash dynamics seen in global food prices \cite{food_prices,Feb_update}. 

The very existence of bubbles and crashes has long been a subject of controversy in economic theory \cite{fama,shiller}, but a direct mathematical formulation of their dynamics has not been framed. The breakdown of equilibrium due to trend-following has been well-established since 1990 \cite{delong}. However, the theory at that point did not represent the bubble and crash dynamics of the system. Why didn't traditional methodologies consider the dynamic effects of interdependencies? As described earlier, a key limitation is that traditional mathematical methods in economics are applicable only to systems in which there is a separation of behavior between the micro and macro scales. Interactions among the parts that cause large scale behavior, like the trend-following-induced market bubbles, violate this separation. Instead, many papers noted the possible role of speculators but did not model their effects mathematically. Recent work on agent modeling demonstrated that market dynamics can deviate from equilibrium prices \cite{lebaron1999time,muchnik2003statistical,lux1999scaling,hommes2008complex} but these approaches were not used for modeling food prices, and the modeling of market agents generally results in a large number of parameters that describe many possible agent behaviors. By including just the largest scale terms in the dynamical expansion of the behavior of prices in Eq. \ref{eq:model0}, and including terms that describe market switching behavior, accurate description of prices was achieved. 

It is also important to recognize the difference between this approach and time series analyses based upon correlations. In such an analysis multiple time series are correlated to a particular target time series, and the correlations are used as a measure of the influence (or at least the association) of the time series to the target time series. Such analyses of correlations do not characterize the scale of those effects, as a correlation is independent of the scale. This, combined with the statistics of correlations, i.e. given a large number of different times series it is a statistical certainty that some of them will be correlated even if they are unrelated, leads to many potentially false associations of quantities that are unrelated to each other. Careful methodologies are needed to ensure that errors do not take place and multiple critiques of statistically based experiments suggest that errors are common in science \cite{staterrors1,staterrors2,staterrors3,staterrors4}. In contrast an analysis of scale can rule out many of the potential relationships, especially at the largest scale of a system. 

\subsection{Network dynamics and collective behaviors: Panic and market prices}

Other examples of application to market prices include a study of panic on the stock market. This analysis began from a universal model of the dynamics of systems having a network of internal influences and external forces \cite{netdyn}. Its validation through application to markets \cite{panic} has been recently published. 

Several strategies have been used to evaluate the behavior of markets from a complex systems perspective. These include network topology models, agent based models, and the approach we are discussing that focuses on large scale behaviors. We therefore provide extensive references in this case to illustrate and contrast different methodologies. We first introduce the essential questions and motivation. 

Panic is one of the important collective behaviors in social systems with relevance to sociology and economics. 
In sociology \cite{wolfenstein,smelser,quarantelli,mawson}, panic has been defined as a collective flight from a real or imagined threat. In economics, bank runs occur at least in part because of the risk to the individual from the bank run itself---and may be triggered by predisposing conditions, external (perhaps catastrophic) events, or even randomly \cite{diamonddybvig,calomirisgorton1990}. 
While market behavior is often considered to reflect external economic news, empirical evidence suggests that external events are not the only 
cause of market panics\cite{cutlerpoterbasummers}. Although empirical studies of panic are difficult \cite{mannnageldowling,galbraith,kindleberger}, efforts to distinguish endogenous (self-generated) and exogenous market panics from oscillations of market indices have met with some success\cite{sornette8,feigenbaum1996,sornette7,stauf,sornette-book}, though the conclusions have been debated\cite{feigenbaum2,sornette4,bree,cho}. 

The 2007-2008 financial crisis led to new interest in analyzing the behavior of market crashes. Whether the 2008 crash resulted from compelling economic news or panic that may or may not have been linked to negative news is an important question. In general, the behavior of markets that fall rapidly then recover shortly afterwards makes suspect the economic assumption that markets reflect valid economic news. However, economic theory does not provide a solid basis for formalizing market behavior outside of the equilibrium concepts in which price is determined by fundamental value. This is not surprising because the ideas of economic equilibrium are the same statistical concepts and mathematical optimization methods that are present in pre-renormalization group thermodynamics and statistical mechanics. On the other hand, panic and other collective social behaviors are to be understood as self-generated patterns of behavior in which interactions between multiple individuals give rise to large scale dynamics without necessitating external forces. These kinds of ordered behaviors, and the transition between disorder and order that might occur in a panic, are precisely the conditions under which the statistical assumptions and optimization based upon averaging assumptions over space (homogeneity across individuals) and time (steady state) do not apply. 

From the perspective of methods based on network topology, the essential approach is to consider a system characterized by sparse networks with heterogenous node connectivities \cite{bararev}, and the dynamics of those connectivities \cite{braha, braha1}. To identify signatures of risk, financial networks \cite{mantegna1, garas, schweitzer, emmert} have been defined primarily from correlations of prices of stocks. Stocks are linked when they have a correlation higher than a threshold, which is placed high enough so that the links are sparse \cite{mantegna, mantegna1, vandewalle, bonnanno, onnela, marketnets, kenett, kenett1}. For example, Bonnanno et al \cite{bonnanno} show that a spanning tree description of correlations \cite{mantegna1} shrinks topologically and has distinct power-law exponents during ``crash" periods. Harmon et al \cite{marketnets} analyze the network to reveal how the crisis propagated between sectors of the economy, suggesting that firewalls would reduce the largest scale risks. 

Agent based models of the trading strategies of market participants are also used to characterize market behavior. Simulations often consider two groups of market participants: `fundamentalists' and `noise traders' \cite{hommes2008complex,lux1999scaling,muchnik2003statistical}. Fundamentalists consider the value of the asset, while noise traders also consider the dynamics of prices, which may result in herding. Simulations suggest \cite{lux1999scaling} that empirically observed volatility clustering and power-law scaling \cite{mantegna,guillaume,gopikrishnan} emerge as traders move from one  group to another. 

A third approach, which we are describing in this paper, the multiscale characterization of complex systems, focuses on modeling the collective dynamics of market prices \cite{chinellato,chinellato1,marketcrises11} and its representation by a minimal set of relevant parameters.  

The multiscale model developed for market crashes focuses on two characterizations of market price behavior. The first is that of traditional economic theory, which considers market prices to reflect perceptions of fundamental value, and therefore changes in market prices to be driven by news, i.e., new information that changes perceptions of fundamental value. The second, of internal self-reinforcing behaviors that can also give rise to price dynamics. Incorporating both, a universal representation can be constructed of the largest scale system behavior when there is both external and mutual influence. The resulting network response model was analyzed to obtain the exact statistical distributions of the fraction of elements that move in the same direction at the same time (the ``co-movement'' fraction) for fully connected networks of arbitrary size \cite{chinellato,chinellato1}. The results are also excellent approximations for other network topologies, including random, regular lattice, scale-free and small-world networks, when normalized to take into account the effect of topology on coupling to the environment. Consistent with the concept of universality, this model and its analytical results can describe a wide variety of networked systems, from Glauber dynamics of the Ising model \cite{chinellato,chinellato1} and evolutionary dynamics in population genetics \cite{ewens,Aguiar2011} to opinion dynamics on social networks reflecting conformity and non-conformity in social systems \cite{arthur}. Here we describe an application to financial markets.

Across the parameter space of the model, the system behavior demonstrates an order-to-disorder phase transition, similar to the water to vapor phase diagram in Fig. \ref{fig3}. There are two parameters of the model, together they control the relative importance of internal and external causes, and the relative proportion of positive and negative external influences. As we move around in the parameter space, three different types of behavior are observed in the statistical distributions of the ``co-movement'' fraction. In the disordered regime where the internal interactions are weak (corresponding to high temperature in the water phase diagram) we have an ``up'' phase corresponding to skewed distributions with a high fraction of stocks that move up (positive price movement), and a ``down'' phase corresponding to skewed distributions with a low fraction of stocks that move up (negative price movement). When the up and down are balanced, the distribution has a single peak at $50\%$ fraction moving up and down. As the strength of interactions between elements increases there is a transition point to collective order. In the ordered regime there are bimodal distributions in which two different phases may exist due to internal interactions causing either mostly up or mostly down movements with slow dynamical switching between them. The switching (flipping) corresponds to a first order phase transition boundary. The second order critical point of this model, the transition between disordered and ordered states, is a unique state with a flat distribution.  

The model is relevant to dynamics of multiple equities, rather than individual stock behaviors. The behavior can still be considered to arise from trading agents, and might be represented by networks of influence between them. However, many of the details are not relevant and are thus abstracted into the aggregate behavior. Thus, for example, as indicated above, the structure of the network does not change the behavior, and unlike trader agent models the behavioral rules of price agents need not differ. The natural behavior that we find is a transition between independent and collective action, the latter of which can be identified with panic. This enables identifying a measure of collective panic and its use to predict financial crises that follow when panic occurs.

Linking concepts of panic to the influence model, mimicry in panic corresponds to mutual influence. The most standard measures of market dynamics are volatility and average correlation. Instead of these, the universal model suggests co-movement as a measure of large scale collective behavior. An essential difference between co-movement and volatility or correlations, is that both volatility and correlations become larger if the same relationships exist among the price movements but the individual price movements become larger. This can lead to having a large volatility or average correlation even when the price movements are mostly independent. On the other hand co-movement may be large even when price movements are small. Thus, co-movement may be a much more direct measure of the collective behavior that we are looking for. It makes sense to hypothesize that co-movement is useful for characterizing panic and nervousness. 

The predictions of the universal model that were solved to give distributions of co-movement were tested empirically against the daily market behavior. Remarkably, the predictions are confirmed on the real-world financial data, covering the recent economic crisis as well as earlier market dynamics. The co-movement data then can be used to evaluate whether the recent market crisis and historical one-day crashes are internally generated or externally triggered. Over the period of the analysis the real world behavior narrowly adopts only the balanced positive and negative news one-dimensional subspace of the parameter space. The critical point with high levels of co-movement, i.e. panic, is found to uniquely identify the 2008 market crash. Since the critical point is unique, no model parameters are adjusted to obtain this correspondence, so this can be considered as a zero parameter theory of the financial crisis. Moreover, a measure of co-movement increases well before one day market crashes, and there is significant advance warning to provide a clear indicator of an impending crash. Increasingly panicky behavior is thus an early warning sign of each market crashes.
Predictive performance is exceptional---it anticipates the largest one-day crashes over 25 years, with no false positives or negatives. Other measures that might be used as predictors of market crises are volatility, correlations and covariance between equity prices. Of these indicators, volatility and correlations, the most commonly used risk predictors, provide the least predictive ability with three errors and four correct predictions, covariance is a comparatively better predictor with only one error, and the multiscale model provides the best predictive utility with no errors \cite{marketcrises11}. 

We note that the distribution of sizes of the co-movement fraction (number of stock prices moving in the same direction, i.e. aligned members) is a multiscale decomposition similar to the complexity profile. 

\subsection{Principles and multiscale analysis}

In this section we review several general features and principles of multiscale analysis as embodied in the complexity profile for use in applications across complex systems. A more extensive discussion of the principles of multiscale analysis and its use in framing a fundamental theory of structure, as well as other measures of multiscale structure (the marginal utility of information), has been provided elsewhere  \cite{allenstacey}. These principles also provide a means of understanding a class of applications to social and biological systems that has been developed to understand the relationship of structure and function across a variety of systems. The development of models using multiscale information requires specific data describing both the scale and dynamical behavior of a system. There are circumstances, however, when the amount of information at a particular scale itself provides insight into the behavior of a system. Here we focus on a few such examples in order to illustrate the utility of the complexity profile. 

\subsection{Sum rule: Elementary excitation strength}

Multiscale analysis evaluates the behavior of systems across scales. The extent or number of degrees of freedom at a particular scale is given by the complexity profile measured in the same way that the entropy describes the degrees of freedom at microscopic scales. It can be shown that there is a fundamental tradeoff between the degrees of freedom across scales \cite{dcs,multiscaleACS,multiscalevariety}. This arises because larger scale behaviors arise due to coherence, i.e. constraints, across microscopic degrees of freedom. The larger the number of components that are involved in a behavior, the larger the scale of that behavior and correspondingly the greater is the reduction in the degrees of freedom at smaller scales. This fundamental property of systems can be framed as a sum rule, i.e. the sum of scale weighted degrees of freedom is independent of the structure of the system but only dependent on the (number of) components of which the system is composed. This sum rule can be seen to be similar to sum rules of the dielectric response function, i.e. that an excitation that is a collective behavior, such as a plasmon, reduces the contribution to response (weight) of single particle elementary excitations \cite{plasmon}. 

\subsection{Ashby's law and multiscale information}

A central tenet in cybernetics is that the number of possible states of a control system---a biological or engineered system that strives to control its environment---has to match the set of possible states of the environment in order for the control system to function successfully. The degree to which even an optimally designed system must fail, is related to the relative number of its available states to the number of states of the environment. This is Ashby's Law of Requisite Variety, which is proven from the assumption that each of the states of the environment must be matched to a different system state in order for the impact of the environment to be mitigated.  

This statement is, however, framed in a context where environmental distinctions are essentially of the same scale and that scale is matched to the differences in state of the system. A microscopic difference in state of the environment need not be met by a difference in the state of a macroscopic control system. Also, a large force exerted by the environment may never be successfully met by a change of state of the system. To incorporate these observations, a multiscale generalization of Ashby's Law is needed in which scale of environmental and system behaviors are recognized. A direct generalization would involve counting the sates of the system at the relevant scale of impact of the environment \cite{multiscalevariety}. 

This directly connects the complexity profile as a measure of multiscale variety with the effectiveness of the behavior of a system. 

\subsection{Formal theorems about complexity profile}

There are theorems proven about the complexity profile that aid in thinking about and approximating it for real world systems. Specifically: 

Independent component property---the complexity profile is the sum over profiles of independent components, X:
\begin{equation}
C_S(\sigma) = \sum_{X} C_X
\end{equation}
Fully dependent component property---the complexity profile of $n$ fully dependent components, $X$, add in scale:
\begin{equation}
C_S(\sigma) = C_X(\sigma/n)
\end{equation}
Combining these together we have that the complexity profile can be given by fully independent and fully dependent components as
\begin{equation}
C_S(\sigma) = \sum_{X} C_X(\sigma/n(X))
\end{equation}
for this to hold, the distinct sets $X$ of $n(X)$ components must be independent of each other. We can consider $n(X)$ to be the scale of sets of components $X$.

\subsection{Calculations of multiscale information based upon data}

There are a number of formulas that have been developed to calculate the complexity profile based upon data about a system behavior. 

The first of these is obtained from the probability of the system across states of the system $P({s_i})$. While this formula has been proven to be unique, it is computationally difficult due to combinatorial treatment of all subsystems, and has been calculated for model systems only for systems of limited sizes including Ising models, Gaussian correlations, and a few other systems \cite{gaussianmultiscale,isingmultiscale,strongemergence}. 

A much more computationally tractable method has been developed to evaluate an approximation to the complexity profile based upon sampled data about the system \cite{yavni}. This method considers the components of a system that are linked to each other and aggregates the number of components that are linked to a certain degree at the scale of that number. 

Even simpler approximations and calculations of the system are possible that are useful for applications to real world problems. In particular, we can approximate the complexity profile by considering sets of components that are behaviorally linked. Counting the components of each set we approximate them as fully connected to each other. Counting the number of sets of a certain size or larger gives the value of the complexity profile at that size. This is an approximation to the complexity profile based upon the theorems of the previous section. This construction can be helpful when there are sufficiently clear distinctions between what constitutes dependent and independent components. Improved approximations consider the degree of dependency as given in Ref. \cite{yavni}.

\subsection{Organizational response}

A first approximation to the complexity profile of a social system is to consider multiple individuals engaging in a particular activity, whose actions are linked to each other as a measure of the scale of that activity. Each activity then is describable as a specific state of those individuals. For example, two people using a crosscut saw must adopt corresponding actions. Multiple two person teams may be working at the same time, but their actions are not necessarily coordinated. The complexity at a particular scale of the system is the sum over the groups (possibly independent individuals) who are engaged in coordinated acts as part of a social organization. This approximation can then be combined with the multiscale law of requisite variety to analyze the performance, or lack of performance, of specific organizations, i.e. social systems organized with the intention of performing particular functions. In this context, we are interested in the behavior across scales, ranging from individual to societal. Despite their wide range of scales, these are all large scale behaviors from the point of view of fundamental physics. 

Thus, we can compare organizations that act based upon individuals performing independent tasks, or large coordinated groups performing a coordinated action. The multiscale version of Ashby's Law now becomes the statement that large scale tasks require large groups acting in a coordinated way, while comparatively small (individual) scale high complexity tasks require independent individuals performing distinct tasks. This is apparent in the contrast between large scale military confrontation---such as during the gulf war, in which hundreds of thousands of individuals moved in a coordinated fashion during the ground troop movements of February 24-28 1991---and normal diagnostic medical services---in which hundreds of thousands of physicians perform medical diagnoses on distinct individuals with diverse conditions. The recognition of the distinction of different types of tasks and the relationship of organizational structures to the need to perform different tasks then serves as a basis for developing principles of organizational structure for management. 

General analyses of the relationship of structure and function can be inferred, specifically the limitation of central control structures for performing highly complex collective tasks \cite{multiscalevariety}. This limitation arises because a hierarchy limits the extent of information that can be communicated among the sub-groups of the hierarchy. 

More specific insights can be obtained \cite{mtw}, these include the recognition that the modern strategy of financial management of medical services runs counter to the ability of the medical system to provide small scale complex care to individuals \cite{ajph}, explaining the fundamental failure to provide high quality services. Similarly, that the education system, through centralization and standardized testing, cannot match its actions to individual differences. In a military context, the ability to respond successfully to highly complex military encounters is not possible for conventional forces but is progressively more effectively met by individually trained marines and even better, special forces \cite{sentinel}. Also, similarly, complex engineering projects cannot be performed by hierarchical decomposition as is described in the conventional waterfall process \cite{eng}. 

These observations can be inferred from the first order characterization of the complexity profile of different systems. They reflect on the overall functioning of social systems that are designed for performing different tasks. More detailed quantitative applications of the complexity profile to such systems is possible. Essential insights into structural ineffectiveness can be gained based upon manifest incompatibilities between the organizational structure and the tasks using multiscale information theory and the  multiscale version of Ashby's Law. 

\section{Summary}

We can expect that developing models of complex systems will require major theoretical and experimental efforts for validation. It is not sufficient to create a model---it has to be analyzed for its ability to capture the behavior of the system and be robust to fine scale details. The difficulty is in recognizing universality classes and the dependencies on relevant parameters. These are technically demanding issues that require substantial care. 

Why don't we just include more details? If we include enough, won't a model be correct? The answer is no, for two reasons. The first is sufficient but the second is more important. The first reason is that including many details without determining what is and is not important cannot tell us whether we have included the details that matter. The second reason is that including many details that don't matter actually prevents us from addressing the question we really want to answer: which levers are important? Determining the levers that are important is equivalent to determining what is important at the larger scales. Thus, the questions we really want to answer about systems are exactly the same as the determination of which variables are relevant. 

Thus, this demanding process should not be avoided. The construction of large scale faithful models enables investments in phenomenology to come to fruition in their ability to address key questions. The effort to create them is therefore essential and worthwhile.

Considering systems in this way, we should recognize that any mathematical model, and indeed any description or characterization, whether from theory or phenomenology, in words, pictures, movies, numbers or equations, is ``valid'' only because of the irrelevance of details. Moreover, such information applies across different instances, because of the sufficiency of the representation having captured the important variables. Any two systems we look at (or the same system at different moments in time or different cases of the same system) are different in detail. If we want to say anything meaningful about a system---meaningful in the sense of scientific replicability or in terms of utility of knowledge---the only description that is important is one that has universality, i.e. is independent of details. There is no utility to information that is only true in a particular instance. Thus, all of scientific inquiry should be understood as an inquiry into universality---the determination of the degree to which information is general or specific.

\section{Appendix: Technical points}

While this paper is not intended as a ``how to" prescription, there are several technical points that are relevant to applications in general. Moreover, the strategy of applications can be framed in a general way. 

One of the keys to applying the analysis of large scale behavior to complex physical, biological and social systems quite generally is to recognize the importance of identifying the distinguishable states of the system at larger scales of observation---and to be consistent about the scale of observation at which we are describing the system, i.e. the level of uncertainty or error we can allow in our description. A technical problem in this regard arises from the tendency to use real numbers that do not identify the number of distinct states they are representing, or the relative size of different variables. This is apparent in the common use of algebraic transformation. For example, we often consider $y$ or $\log(y)$ or $e^y$ or $1/y$ to be equally valid variables. However, such transformations change the size of uncertainty of variations at different values of the variable. Thus, using real numbers without specifying the uncertainty and how it depends on value of the variable, i.e. enumerating the distinguishable values of the variable, obscures what we are trying to capture.

The second key is to understand how aggregation works. As we increase the scale, or equivalently, reduce the level of resolution of our observations of a system, we see fewer details. Smaller distinctions disappear and only larger distinctions that involve many parts of the system together remain. How properties aggregate determines what is observed, i.e. what is important. Using the inherent way that things aggregate enables us to figure out what properties of a system are important at the larger scale. Aggregation hinges on how the parts depend on each other. The simplest case is when they are either completely dependent or independent; in this case, aggregation is the averaging we know from statistics that gives rise to a normal (Gaussian) distribution. When elements are dependent, due to being influenced by the same thing, or due to mutual interactions, then the aggregate of $N$ elements scales as the system size $N$. When they are completely independent then the aggregate scales as the square-root of the system size $\sqrt{N}$. The key to understanding other kinds of systems and their scaling behavior is that the dependencies between the parts due to interactions, constraints, and dynamics, give rise to different scaling behaviors.

The third key is recognizing that there is a universal way to represent the behavior of systems. Any system can be decomposed into components and an important way of constructing a model is developing an understanding of how the behavior of the components in aggregate comprise the behavior of the whole if their dependencies are properly accounted for. The property that is important to know about a component is the set of distinguishable states it has. Similarly, the property that is important to know about the system is the set of distinguishable states it has. In addition, we need to know how external influences couple to these states. Thus, the process for developing a model is to: (a) identify the set of elements of a system to be described; (b) identify the set of distinguishable states they have; (c) identify the dependencies among these states; (d) analyze the distinguishable states of the entire system as they arise from the distinguishable states of the components and their dependencies; and (e) characterize the external influence on the set of distinguishable states of the system. 

Because of the pervasive use of calculus and statistics in the analysis of systems today, we must watch out for the inappropriate use of approximations that use those approaches. Among these approximations are ad-hoc assumptions about which variables are the right variables to describe a system, and the explicit or implicit use of averaging and smoothness in formulating the questions to be analyzed. Such assumptions should be considered as approximations, like the mean-field approximation which assumes components are all subject to the same environmental conditions. This approximation, which returns us to the smoothness and independence of calculus and statistics, is often considered to be valid, but is not when patterns of behavior, in space or in time, arise in a system that lead to differences in local context and a need for additional variables. At the same time, we must not make the error of considering the notion that ``all data'' should be captured about a system, or identifying a-priori the description of the system to be used without performing an analysis of which information is really important. Any such effort is incomplete as it is constrained by the nature of the information that is available.

Consider biological examples of what can be distinguished: cell states, organ states, physiological states. Any model must start by identifying which are the states that are distinguishable. This should be done over a macroscopic range of scales that are relevant to observation. This is not easy. Still, it is easier than the problem of describing all of the fine scale details. To identify the behavior of a tissue, we identify the distinguishable states of the cells and the interactions between these distinguishable states. We do not identify all of the details of cell function, only states that are distinguishable. Similarly, to model human beings in society, we only identify human behavioral states that are distinguishable. There may be many reasons for specific individual states, but those reasons are not relevant unless we are considering a person in detail. This, therefore, is an entirely general framework for modeling the behavior of complex systems. 

Ultimately, however, the crux of the analysis that is to be performed is to identify the scale of information: how many components are doing the same thing. The scientific question is to understand what controls how many are doing the same thing. Among the issues that are essential for this determination is identifying: 
\begin{itemize}
\item External forces that act on the system, their strength and the way they influence the system. Barriers/boundaries that limit the influence can play an important role, as well as the identification of which components are directly affected. 
\item Internal mechanisms by which influence, contagion, mimicry, amplification as well as recovery affect the growth of the number of components that are doing the same thing, or coupled things. The way components are connected to each other, across one, two or three spatial dimensions, or more generally in a network, often matters. Also, influences among components can be simply reinforcing, or cause opposing or other relationships that are behaviorally coupled. 
\item Noise and its role in triggering new behaviors, as well as independence and dissipation. Noise can be considered as a kind of external force that satisfies the traditional statistical assumptions associated with fine scale behaviors. Individual events are often approximated as having a localized and small scale influence, and independent events are often considered to be pervasive across the system, affecting all components directly.
\end{itemize}
Isolated events arising from large scale external forces have a trajectory of growth in size that is a cascade of effects. The accumulation of the effects of noise events are often key to power-law (fractal) scaling behaviors. Spatial patterns occur when influences are reinforcing over some distance and anti-reinforcing over a larger distance. Oscillations occur when the same happens over time. Identifying the specific behavior that occurs then becomes a task in characterizing only a few key aspects of the external forces, internal mechanisms, and noise. 

\section{Acknowledgements}

Many thanks to Maya Bialik for collaboration on an earlier version \cite{bialik}, and to Irving Epstein, Matthew Hardcastle, Karla Z. Bertrand, Blake Stacey, Casey Friedman and Dominic Albino for helpful comments.

\end{document}